# AUTO ENCODER CONVOLUTIONAL NEURAL NETWORK FOR PNEUMONIA DETECTION

Michael Nosa-Omoruyi and Linda U. Oghenekaro

Department of Computer Science, University of Port Harcourt, Port Harcourt, Nigeria

## ABSTRACT

*This study presents an innovative approach utilising Autoencoder Convolutional Neural Networks (AE-CNNs) for pneumonia detection in paediatric chest x-rays. The research addresses the complexity of pneumonia, considering diverse causative agents, including bacteria, viruses, and aspiration. Autoencoder Convolutional Neural Networks are employed to enhance anomaly detection by revealing hidden patterns in the data. The evaluation process involves meticulous analysis of the histogram reconstruction error, leading to the establishment of a threshold for anomaly identification. The results demonstrate distinct differences in error magnitudes during testing and training periods, with a threshold providing a tangible criterion for anomaly detection. The study contributes valuable insights into the discriminative capability of Autoencoder Convolutional Neural Networks, with a threshold of 0.0127, in detecting pneumonia in paediatric chest x-rays, emphasising their potential for improving diagnostic precision.*

## KEYWORDS

*Convolutional Neural Network, AutoEncoder, Pneumonia, Chest X-rays, Anomaly Detection*

## 1. INTRODUCTION

Medical imaging [1], [2] has developed as a critical and transformational component of modern healthcare, altering how healthcare practitioners approach the diagnosis, treatment, and monitoring of medical disorders. This evolution has been fuelled by constant technical advancements, resulting in a wide range of imaging modalities and techniques, each suited to fulfil the varied demands of medical practice. The non-invasive nature of medical imaging is a fundamental and differentiating feature, eliminating the need for invasive treatments and surgical interventions to appreciate the complexity of the interior constituents of the human body. This non-invasiveness is notable for its patient-centric characteristics, which alleviate discomfort and expedite recovery times while simultaneously improving the accuracy and expeditiousness of diagnosis, resulting in improved patient outcomes. Central to medical imaging is its capacity to offer insights into the human body hitherto beyond reach. These images can be likened to a portal into the systems of human anatomy, endowing healthcare practitioners with a visual conduit to explore, analyze, and pinpoint structural anomalies, functional intricacies, and pathological aberrations. From the identification of fractures and neoplastic growths to the monitoring of embryonic development and the evaluation of cardiac performance, the applications of medical imaging span a broad spectrum of medical disciplines. Medical imaging has an impact that reaches beyond patient treatment. It plays a pivotal role in medical research, medical pedagogy, and healthcare practitioner professional development. The massive amount of data created by these technologies helps significantly to scientific discoveries, spurring the development of ground-breaking treatments and diagnostic procedures. It also functions as an effective teaching tool, allowing healthcare professionals to broaden their knowledge and skill in their respective fields. In terms of the future, the trajectory of medical imaging appears promising. Consistent technological advances, particularly the incorporation of artificial intelligence (AI), promise to further revolutionize the field.

AI promises to expedite the interpretation of medical images, automate image analysis, and foster heightened diagnostic precision, thereby streamlining the diagnostic process. Anomaly detection is a critical activity in a variety of industries, including manufacturing, cybersecurity, healthcare, and video surveillance. Its primary goal is to identify unusual patterns or events that deviate significantly from established norms within a given dataset. Traditional methods for addressing this difficulty have frequently struggled with the complexities inherent in high-dimensional data and elaborate patterns.

Pneumonia [3]–[5], a global health concern characterized by inflammation affecting lung alveoli, stems from diverse origins, including bacteria, viruses, fungi, and aspiration. Common culprits like Streptococcus pneumoniae and influenza viruses manifest through symptoms such as fever, coughing, and shortness of breath. Diagnosis involves a comprehensive clinical evaluation, incorporating medical history, physical examination, radiological imaging, and laboratory tests. Chest X-rays and blood tests aid in identifying infiltrates and specific pathogens, guiding targeted treatment strategies. Antibiotics address bacterial pneumonia, while antivirals combat viral cases, with supportive measures like oxygen therapy and fluid management providing comprehensive care. Prevention through vaccines targeting pneumococcal and influenza infections remains crucial. The diverse array of causative agents contributing to pediatric pneumonia underscores the complexity of the condition. Bacterial pneumonia, a prevalent form, is frequently attributed to Streptococcus pneumoniae, Haemophilus influenzae. Each of these bacteria has distinct characteristics, and their prevalence varies based on age groups and environmental factors. Streptococcus pneumoniae [6], commonly referred to as pneumococcus, is a leading cause of bacterial pneumonia in children. This Gram-positive bacterium is known for its ability to colonize the upper respiratory tract, leading to infections when the immune system is compromised or when the pathogen gains access to the lower respiratory tract. Haemophilus influenzae [7], [8], particularly the nontypeable strains, is another prominent bacterial culprit in pediatric pneumonia, especially in children under 5 years old. This bacterium has a propensity to cause respiratory infections and can contribute to the development of pneumonia, often in conjunction with other pathogens. Viral infections also significantly contribute to pediatric pneumonia [9]–[11]. Respiratory Syncytial Virus (RSV) [12]–[14], a leading cause of lower respiratory tract infections in infants and young children, is particularly notorious. RSV infections can progress to pneumonia, especially in vulnerable populations such as premature infants or those with underlying respiratory conditions. Influenza, caused by the influenza virus, is another viral agent capable of inducing severe pneumonia in children. During seasonal outbreaks, the influenza virus can lead to a surge in pediatric pneumonia cases, necessitating heightened vigilance during these periods. Aspiration pneumonia [15]– [17] adds another layer of complexity to the spectrum of causes. This form of pneumonia results from the inhalation of foreign substances, such as food particles, liquids, or vomited material, into the lungs. Aspiration can occur in various settings, including during feeding or in individuals with impaired swallowing reflexes, leading to inflammation and infection in the lungs. In the field of artificial intelligence, Convolutional Neural Networks (CNNs) [18]–[20] mark a significant breakthrough, particularly in computer vision. Building on neural network principles, Convolutional Neural Networks utilize neurons, layers, activation functions, and backpropagation. Convolutional layers employ filters and kernels for feature extraction, while pooling layers spatially reduce dimensions. The architectural framework includes input layers, convolutional layers for feature extraction, activation functions for non-linearity, pooling layers for spatial reduction, fully connected layers for abstraction, and output layers for predictions. Functionally, CNNs excel in hierarchical feature learning through supervised training, incorporating loss functions, optimization algorithms, and regularization techniques. Autoencoders [21]–[23], featuring an encoder-decoder architecture for unsupervised learning, compress input data into a latent representation reconstructed by the decoder, with training focused on reducing reconstruction errors. Types of autoencoders include sparse, denoising, and variational autoencoders, serving applications like data compression and anomaly detection. Autoencoder Convolutional Neural Networks (AE-CNNs) [24] have emerged as a robust solution for addressing challenges in medical imaging, particularly in pneumonia detection. These models enhance prognosis by revealing hidden facts in the data. The anomaly identification process involves several stages: training, where an autoencoder is set up with hyperparameters; testing, where the autoencoder reconstructs input data from the latent space representation with minimal variance; evaluation, comparing Mean Squared Error (MSE) values of training and test datasets; and validation, using efficiency metrics like accuracy and MSE scores to discern anomalies. This innovative approach enables the development of adaptable and accurate learning models for detecting abnormalities in medical imaging.

## 2. RELATED WORK

Siddalingappa, R.et al [25] study explores anomaly detection in medical image datasets, focusing on autoencoders and convolutional neural networks (CNNs). The approach involves training an autoencoder on lung cancer CT scan images, assessing anomalies through Mean Squared Error (MSE) during testing, and validating with accuracy metrics. The method achieves high accuracy (98% for outlier detection, 97.2% for classification), demonstrating its efficacy in isolating anomalies within lung cancer CT scans. This work contributes valuable insights to the broader field of anomaly detection in medical imaging, showcasing the potential of autoencoder CNN integration for improved diagnostic capabilities.

Ruixuan et al [26] explores in chest radiography for medical screenings, a novel abnormality detection method is introduced, tailored for healthy X-ray images and employing an autoencoder. This model reconstructs the normal image while predicting pixel-wise uncertainty. Higher uncertainty at normal region boundaries, indicative of larger reconstruction errors, is a key observation. This informs the metric: normalizing the reconstruction error by uncertainty for natural abnormality measurement. Rigorous testing on two chest X-ray datasets establishes the method as a state-of-the-art solution for autonomous abnormality detection.

Owayed, A. F., et al [27] study aims to describe underlying illnesses in children with recurrent pneumonia hospitalized in a tertiary care pediatric hospital. The goal is to propose a series of investigations for children experiencing recurrent pneumonia, addressing gaps in Pneumonia is a major issue in children, causing up to 5 million deaths annually in developing countries. In North America, annual pneumonia incidence ranges from 30 to 45 cases per 1000 children under 5 years and 16 to 22 cases per 1000 children aged 5 and older. Recurrent pneumonia, defined as at least 2 episodes in a year or more than 3 at any time, poses a concern.

Farha, T. et al [28] studied that in developed nations, comprehensive epidemiological studies on pneumonia are limited, emphasizing the importance of accurate ascertainment and definition for estimating incidence in primary care and hospitals. The data indicates an annual burden of 10–15 cases per 1000 children, with hospital admission rates of 1–4 per 1000, peaking in the youngest children and decreasing after age 5. Most cases of community-acquired pneumonia lack identified organisms, with common viral culprits including influenza and respiratory syncytial virus. Streptococcus pneumoniae is a prevalent bacterial cause. The estimated annual cost of childhood pneumonia to the NHS in England is £6–£8 million, excluding family and social costs. Innovative vaccine strategies hold promise for reducing both incidence and economic burdens.
Ruuskanen, O., et al [29] introduced that every year, there are approximately 200 million cases of viral community acquired pneumonia, affecting both children and adults in equal measure.

Molecular diagnostic advancements have uncovered a higher incidence than previously believed. In children, the culprits often include respiratory syncytial virus, rhinovirus, human metapneumovirus, human bocavirus, and parainfluenza viruses. Notably, dual viral infections and coinfections with bacteria are common in this demographic. For adults, viruses are responsible for about one-third of community-acquired pneumonia cases, with influenza, rhinoviruses, and coronaviruses being prevalent. However, bacterial infections continue to be predominant in the adult population. Distinguishing between viral and bacterial pneumonia relies on various factors, such as age, symptom onset, biomarkers, radiographic changes, and response to treatment. Despite these considerations, a definitive clinical algorithm is yet to be established, and there is a lack of consensus regarding antibiotic use for viral pneumonia.

De Benedictis, F. M., et al [30] introduced a study on complicated community-acquired pneumonia in previously healthy children is a severe illness with local (e.g., effusion, empyema) and systemic complications (e.g., bacteraemia, multiorgan failure). Common pathogens are Streptococcus pneumoniae and Staphylococcus aureus. Diagnosis involves chest radiography and ultrasound, excluding routine CT scans. Treatment includes prolonged intravenous antibiotics, guided by local microbiology and culture results. Drainage may be needed, with rare use of intrapleural fibrinolytics. Extensive surgery is seldom required, especially in resource-limited settings. The clinical course can be prolonged, but complete recovery is the usual outcome, even with necrotising pneumonia.

Crame, E., et al [31] discusses pneumonia acquired within the community which remains a leading cause of mortality among children under the age of 5 globally. The etiology of community-acquired pneumonia encompasses bacterial, viral, and fungal origins, presenting a clinical challenge in distinguishing between them. Although viral pathogens are frequently identified as the primary cause, bacterial infections are associated with more severe manifestations. It is crucial for clinicians to skilfully discern cases requiring additional treatment or hospital admission from those manageable at home. Given the diverse array of symptoms and potential complications, pediatricians face a complex task in addressing pneumonia in children.

Field, E. L., et al [32] introduces a systematic review of existing literature to consolidate and summarize the evidence concerning the effectiveness of artificial intelligence (AI) in the classification of pediatric pneumonia based on chest radiographs (CXRs). After an initial search for studies meeting predefined criteria, data extraction was carried out using a dedicated tool, and the selected studies underwent assessment using critical appraisal tools to evaluate the risk of bias. The gathered results were analyzed, with sensitivity, specificity, accuracy, and area under the curve (AUC) as the key outcome measures. Five studies met the inclusion criteria. The ensemble AI algorithm demonstrated the highest sensitivity at 96.3%, while DenseNet201 achieved the highest specificity (94%) and accuracy (95%). The VGG16 algorithm exhibited the most notable AUC value at 96.2%. Notably, certain AI models approached a diagnostic accuracy of nearly 100%. To gauge the efficacy of AI in a clinical context, a comparison with radiologists' performance is crucial. The evaluated AI algorithms displayed promising results, suggesting their potential to streamline and expedite diagnoses once replicated studies validate their performance in clinical settings. This could potentially lead to significant advancements in saving lives on a large scale.

## 3. METHODOLOGY

A diverse dataset of pediatric chest x-ray images was collected, ensuring representation from both normal and anomalous cases. The images underwent a meticulous preprocessing phase, including resizing, normalisation, and augmentation, to enhance the quality of the dataset and diversity. The initial phase focused on developing an autoencoder architecture to capture intrinsic features from normal chest x-rays. The AutoEncoder was trained iteratively, with hyperparameter adjustments based on validation performance. Encoded features derived from the trained AutoEncoder were analysed to ensure meaningful representation of normalcy in paediatric chest Xrays. This iterative process allowed for a thorough understanding of the relevance of the features in subsequent stages of the study. A Convolutional Neural Network (CNN) architecture was designed for classifying normal and abnormal chest X-rays. The encoded features from the AutoEncoder were seamlessly integrated into the Convolutional Neural Network (CNN) model. Iterative training included considerations for both reconstruction and classification losses, with architecture refinement guided by ongoing validation and expert feedback. The AutoEncoder and components were combined into an Autoencoder Convolutional Neural Network (AE-CNN) model. Joint training sessions were conducted to optimise the overall model, fine-tuning hyperparameters based on continuous validation feedback. The AutoEncoder Convolutional Neural Network model underwent comprehensive evaluation on a dedicated test set, employing metrics such as reconstruction error. The anomaly detection threshold was used to strike an optimal difference between normal and abnormal samples.

### 3.1. Use Case Diagram of the Autoencoder Convolutional Neural Network

This use case diagram in Figure 1, outlines the seamless interaction between the user and the Autoencoder Convolutional Neural Network (AE-CNN) system for anomaly detection in pediatric chest X-rays. The process begins with the user uploading a pediatric chest X-ray image for analysis. This initial step sets the stage for the subsequent anomaly detection process. Following the image upload, the user triggers the anomaly detection process. By doing so, the user instructs the Autoencoder Convolutional Neural Network system to meticulously analyse the X-ray image for potential anomalies. The Autoencoder Convolutional Neural Network system, equipped with Autoencoder Convolutional Neural Network architecture, takes charge of processing the uploaded paediatric chest Xray. This involves the network leveraging its learned features to identify patterns indicative of anomalies. Once the anomaly detection process is completed, the Autoencoder Convolutional Neural Network system provides clear and interpretable results to the user. This may involve highlighting specific regions of the X-ray image that exhibit anomalies. The user gains valuable insights into potential issues, aiding in timely and informed decision-making.

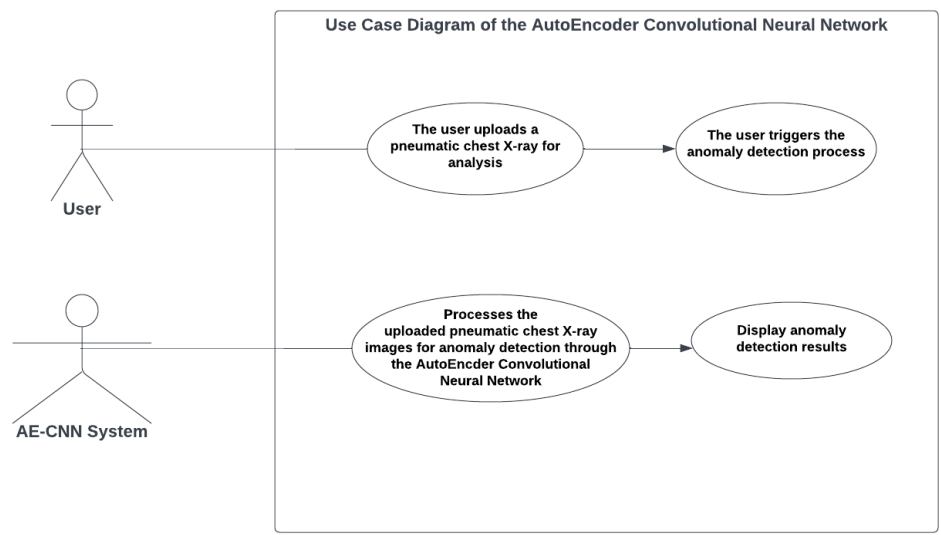

Figure 1. Use Case Diagram of the AutoEncoder Convolutional Neural Network

## 3.2. Activity Diagram of the Autoencoder Convolutional Neural Network

This intricate process involves a collaboration between user-initiated actions and the sophisticated AutoEncoder Convolutional Neural Network system. The activity diagram in Figure 2, seamlessly integrates these components, elucidating the workflow from the commencement of the operation to the conclusive result. The inception of the anomaly detection process is marked by the Start Operation node, symbolising the pivotal role of the user in initiating the AutoEncoder Convolutional Neural Network system. The subsequent step involves the user uploading a chest X-ray image to the AutoEncoder Convolutional Neural Network, signifying the transfer of data from the user to the system for analysis. As the uploaded image enters the system, the journey begins with preprocessing, a crucial phase involving normalisation and resizing. This step ensures that the image is optimised for subsequent analyses, setting the stage for the application of the AutoEncoder Module. This module, equipped with an attention mechanism, focuses on pertinent features within the chest X-ray, amplifying regions that may harbour anomalies. The Convolutional Neural Network (CNN) takes centre stage for feature extraction, unravelling intricate patterns embedded within the X-ray image. This stage harnesses the power of deep learning to capture meaningful representations, laying the groundwork for anomaly detection. Anomaly detection, embodied in the decision process, marks a critical juncture in the functionality of the AutoEncoder Convolutional Neural Network. This stage encompasses the setting of a threshold for anomaly classification and subsequent classification based on the output of the Convolutional Neural Network. It is the culmination of the analytical prowess of the system, where decisions are made regarding the presence or absence of anomalies within the chest X-ray. Post-processing follows suit, refining the results and offering a comprehensive visualisation of detected anomalies. This stage not only enhances the interpretability of the results but also provides a crucial feedback loop for the continuous improvement of the system. The conclusive stage, Display Result, serves as the interface between the system and the user, presenting the final output indicating whether the chest X-ray is deemed normal or exhibits anomalies. The End Operation node signifies the closure of the process, with the user receiving the result of the anomaly detection operation.

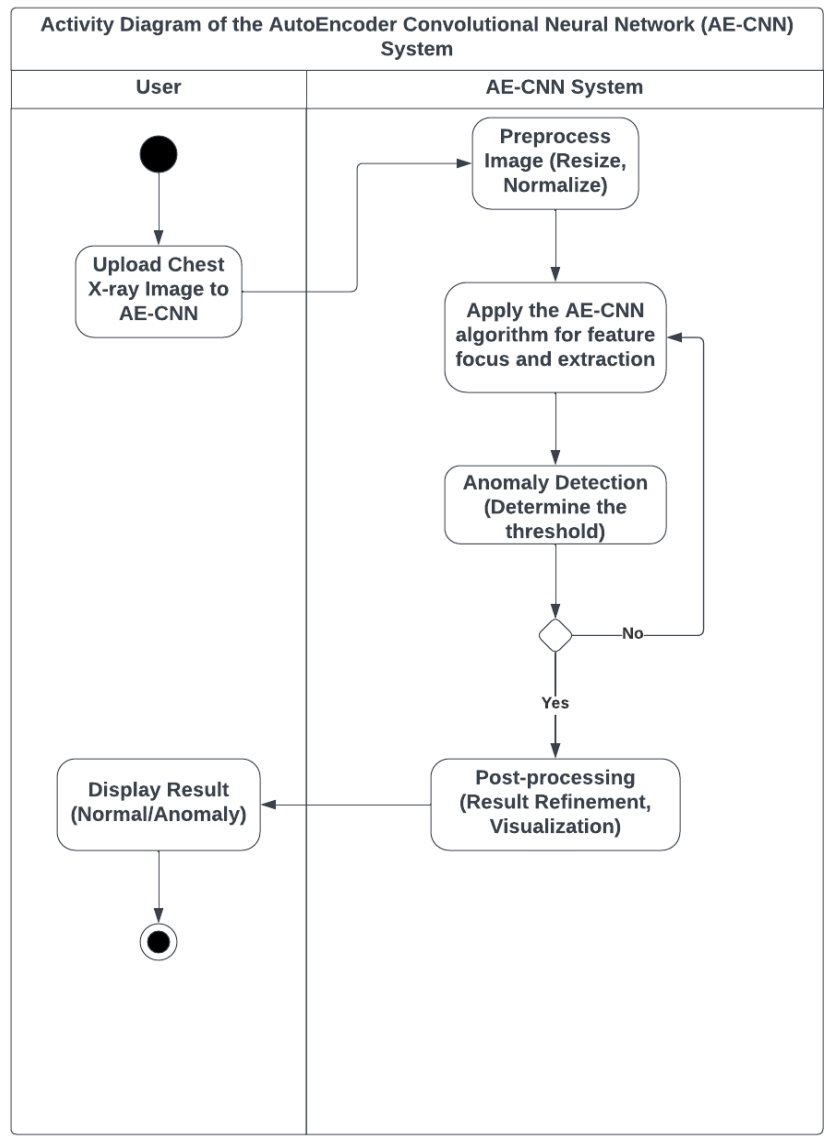

Figure 2. Activity Diagram of the AutoEncoder Convolutional Neural Network

## 3.3. Component Diagram of the Autoencoder Convolutional Neural Network

Figure 3 depicts the architecture of the Autoencoder Convolutional Neural Network (AECNN) system developed for pneumatic detection in pediatric chest X-rays. This model consists of several interconnected modules, each with a critical role in the overall anomaly detection process. The user is central to the system, interacting with the various modules to input data, initiate processes, and receive anomaly detection results. The system's architecture is modular, which means that each component can operate independently while still contributing to overall functionality. This modularity not only makes the system more scalable and maintainable, but it also enables individual modules to be updated and refined independently.

The user first interacts with the Input Module. This component is in charge of receiving data from the user and other external sources. Data may take the form of images. When the data is received,

27

the Input Module preprocesses it to ensure that it is in the proper format for further analysis. This preprocessing step includes normalization and resizing the data in order to prepare it for the next stage of the pipeline.

The data is then fed into the AutoEncoder Convolutional Neural Network (CNN) Module, which serves as the primary system anomaly detection component. An AutoEncoder is a type of neural network that learns a compressed representation of input data, typically for dimensionality reduction or noise removal. In the context of anomaly detection, the AutoEncoder is trained to reconstruct input data as precisely as possible. Because of its ability to capture spatial hierarchies via convolutional layers, the AutoEncoder CNN component is especially well-suited for processing high-dimensional data such as images.

The output of the AutoEncoder CNN is then sent to the Anomaly Detection Module. This module compares the reconstructed data to the original input to identify deviations that may indicate anomalies. The underlying assumption is that the AutoEncoder, which was trained on normal data, will struggle to accurately reconstruct data that deviates significantly from the norm. As a result, any significant reconstruction error can be identified as an anomaly. This module is critical for identifying potential issues that require further investigation.

The system includes a Training Module to ensure that the AutoEncoder CNN detects anomalies accurately. This component oversees training the AutoEncoder on labeled data, which typically consists of a large set of normal data. During training, the model's parameters are adjusted to reduce reconstruction error on normal data while ensuring that it copes with anomalous data. The training process is iterative, with the model being fine-tuned as more data becomes available or as the definition of normal behavior shifts over time.

The Evaluation Module is then used to assess the trained performance of the model. To measure the Area Under Curve (AUC) of the model, this module uses a separate dataset that was not used during training. The Evaluation Module results provide critical feedback on the effectiveness of the model, which may prompt additional training or adjustments if necessary. By continuously evaluating the model, the system ensures that it remains reliable and effective in detecting anomalies over time.

Finally, the user interacts with the system through the User Interface Module, which serves as the bridge between the user and the underlying processes. This module allows the user to input data, configure system parameters, and visualize the results of the anomaly detection process.

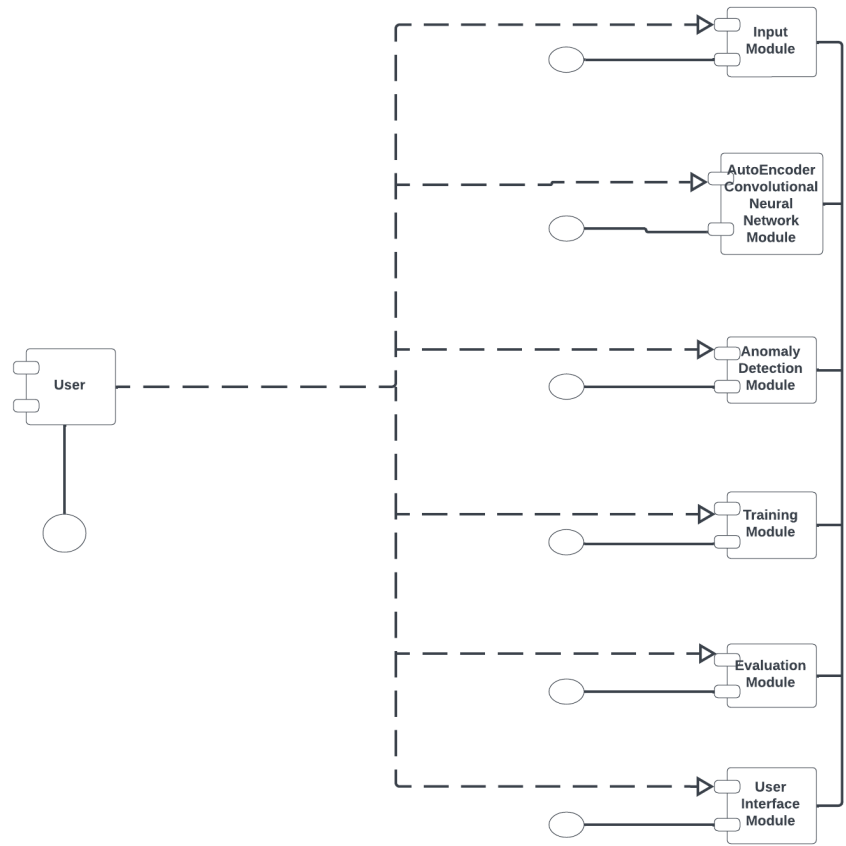

Figure 3. Component Diagram of the AutoEncoder Neural Network

## 4. EVALUATION

The histogram in Figure 4, reveals a well-defined peak in the lower range, suggesting successful reconstruction of normal and anomalous instances. The threshold of 0.0127 serves as a cutoff point, distinguishing between instances with low reconstruction errors (considered normal) and those with higher errors (potentially anomalous). Instances with reconstruction errors above the threshold of 0.0127 are classified as normal, while those below are potential anomalies. The shape of the histogram with a pronounced peak and a spread of errors beyond the threshold, suggests a reasonable ability of the model to discriminate between the two classes.

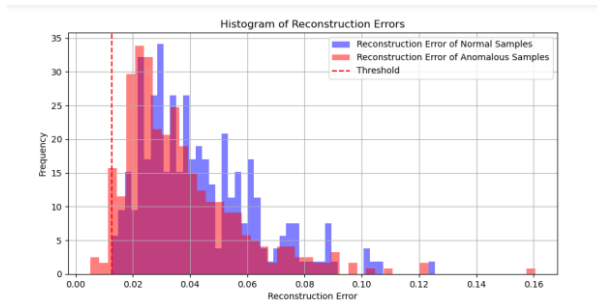

Figure 4. Histogram of Reconstruction Errors with a threshold of 0.0127

## 5. RESULT DISCUSSION

The examination of the performance of the AutoEncoder Convolutional Neural Network (AE-CNN) for pneumonia detection in pediatric chest x-rays reveals distinct differences in error magnitudes between the testing and training periods. Of particular significance is the utilisation of a threshold set at 0.0127, extracted from the reconstruction error histogram depicted in Figure 4. This threshold emerges as a crucial metric, providing a clear basis for identifying anomalous patterns within the data. Additionally, the Area Under Curve (AUC) value of 0.55 offers insights into the discriminatory ability of the model. The chosen threshold of 0.0127, derived from careful analysis of the reconstruction error histogram in Figure 4, serves as a key reference point for anomaly detection. The Area under Curve (AUC) value of 0.55 further contributes to our understanding of the performance of the AutoEncoder Convolutional Neural Network. While not exceptionally high, this metric signifies the model's ability to distinguish between normal and anomalous patterns in paediatric chest X-rays. AUC values closer to 1 indicate a more effective discriminatory model, and in this case, the value of 0.55 suggests a moderate level of discriminative capability. In real-world scenarios, the reconstruction error at any given time emerges as a vital indicator for continuous monitoring. The established threshold of 0.0127 serves as a trigger, prompting the system to automatically generate alerts or emails to users when the reconstruction error exceeds this predefined limit. This proactive approach enables timely interventions, allowing users to address potential anomalies before they escalate.

### 5.1. Performance Evaluation Criteria

The proposed Autoencoder Convolutional Neural Networks (AE-CNNs) were evaluated for pneumonia detection in pediatric chest X-rays using the Histogram of Reconstruction Error as the primary criterion. In the context of AE-CNNs, each input image is encoded and then decoded, and the difference between the original image and its reconstruction is calculated as the reconstruction error. This error indicates how well the model learned to represent the normal data distribution. By plotting these errors in a histogram, we can see the distribution and frequency of various error magnitudes. The histogram enables an empirical approach to determining a threshold for distinguishing between normal and pneumonia cases. By examining where the reconstruction errors diverge significantly, a threshold can be established to maximize both.

## 6. CONCLUSION

This paper represents a comprehensive effort to develop an effective pneumonia detection system for pediatric chest X-rays. By integrating advanced machine learning techniques, iterative development, the resulting AutoEncoder Convolutional Neural Network model demonstrates the potential to enhance clinical workflows and contribute to improved patient outcomes. The flexibility in deployment platforms ensures adaptability to diverse healthcare environments, emphasising the commitment to addressing the unique challenges of anomaly detection in medical imaging.

## AUTHORS

**Michael Nosa-Omoruyi** is a promising academic and researcher in the field of computer science. Currently, he is actively pursuing his bachelor's degree in computer science (in view), showcasing a strong commitment to academic excellence and a passion for advancing technology. His academic journey is marked by a dedication to staying at the forefront of cutting-edge developments in the field.

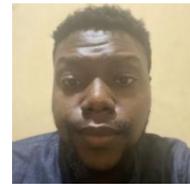

Drawing inspiration from a diverse educational background, Michael is cultivating his expertise in computer science. As he navigates through his academic pursuits, he is actively engaged in research that contributes to the intersection of computer science and industrial applications.

**Dr. Linda Oghenekaro** is a senior lecturer and a seasoned researcher with over 12 years of experience in both industry and academia. She possesses expert technical, programming, and research skills. She is currently the Ag. Head, of the Department of Cybersecurity, at the University of Port Harcourt, Rivers State, Nigeria. Her research interests are machine learning and cyber security.

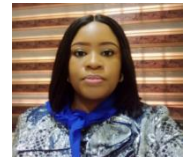

Her publications in both local, and international peer-reviewed journals center around machine learning, and other artificial intelligence solutions, where she has demonstrated the process of developing machine learning models for both classification and categorization problems. She is also actively involved in facilitating software boot camps for the Girl Child, in the bid to build their confidence in delving into STEM disciplines. She is a strong voice in advocating for responsible digital citizenship among the youth. Dr. Linda is an active member of several professional bodies and engages in countless number of community services.